\documentstyle[12pt,amssymb]{article}
\begin{document}

\tolerance=5000

\def\pp{{\, \mid \hskip -1.5mm =}}
\def\cL{{\cal L}}
\def\be{\begin{equation}}
\def\ee{\end{equation}}
\def\bea{\begin{eqnarray}}
\def\eea{\end{eqnarray}}
\def\tr{{\rm tr}\, }
\def\nn{\nonumber \\}
\def\e{{\rm e}}

\begin{titlepage}

\begin{center}
\Large
{\bf Quantum effects and stability of chameleon cosmology}

\vfill

\normalsize

\large{ 
Shin'ichi Nojiri$^\spadesuit$\footnote{Electronic mail: nojiri@nda.ac.jp, 
snojiri@yukawa.kyoto-u.ac.jp} and 
Sergei D. Odintsov$^{\heartsuit\clubsuit}$\footnote{Electronic mail:
 odintsov@ieec.fcr.es Also at TSPU, Tomsk, Russia}}

\normalsize

\vfill

{\em $\spadesuit$ Department of Applied Physics, 
National Defence Academy, \\
Hashirimizu Yokosuka 239-8686, JAPAN}

\ 

{\em $\heartsuit$ Institut d'Estudis Espacials de Catalunya (IEEC), \\
Edifici Nexus, Gran Capit\`a 2-4, 08034 Barcelona, SPAIN}

\ 

{\em $\clubsuit$ Instituci\`o Catalana de Recerca i Estudis 
Avan\c{c}ats (ICREA), Barcelona, SPAIN}

\end{center}

\vfill 

\baselineskip=24pt
\begin{abstract}
One possibility to explain the current accelerated expansion of the universe
may be related with the presence of cosmologically evolving scalar 
whose mass depends on the local matter density (chameleon cosmology).
We point out that matter quantum effects in such scalar-tensor theory 
produce the chameleon scalar field dependent conformal anomaly.
Such conformal anomaly adds higher derivative terms to chameleon 
field equation of motion. As a result, the principal possibility
for instabilities appears. These instabilities seem to be irrelevant 
at small curvature but may become dangerous in the regions
where gravitational field is strong.

\end{abstract}

\noindent
PACS numbers: 98.80.-k,04.50.+h,11.10.Kk,11.10.Wx

\end{titlepage}

The recent astrophysical data \cite{R,B} indicate that 
there is a dark energy providing approximately two thirds of the current universe 
energy density. 
There are various scenarios to explain what the dark energy is.
For instance, the dark energy can be regarded as the (effective) 
cosmological constant which provides the current cosmic acceleration.
Immediately, the question appears: why the cosmological constant, or
accelerating Hubble
constant 
is so small (of the order of $10^{-33}$ eV), compared with the Planck
scale $10^{19}$ GeV?
In another scenario the dark energy is produced by some 
exotic matter like phantom (field with negative kinetic
energy) \cite{caldwell} or some other (usually scalar) matter.
  Unfortunately, such scalar fields are usually
very light.  Its coupling to matter should be tuned to 
extremely small values in order not to be in conflict with the Equivalence 
Principle. In a sense, the cosmological evolution of scalars
contradicts with
the solar system tests.

Recently, the very interesting attempt to overcome the problems with light
scalars has been suggested in \cite{KW} (Chameleon Cosmology, see also \cite{Mota}).  
The effective mass of the chameleon scalar field depends on the local  
matter density
 (for earlier discussions of density-dependent potentials, see
\cite{potentials}). 
Then on the cosmological scales, the evolving chameleon scalar is almost
massless and 
gives naturally an effective cosmological constant of the order of the matter 
density in the universe. The mass becomes heavy, say on the earth. That
does not 
contradict with the solar system tests of the Equivalence Principle. 
The coupling of chameleon with the matter is of the order unity \cite{KW}.
Furthermore the potential and the couplings of the chameleon field(s) 
with matter
are rather usual 
in the string theory. 

We start with quick review of chameleon field scenario.
The initial action   has the
following form:
\bea
\label{I}
S&=&\int d^4 x\sqrt{-g}\left[{M_{\rm Pl}^2 \over 2}R - {1 \over 2}\partial_\sigma \phi 
\partial^\sigma \phi - V(\phi)\right] \nn
&& - \int d^4 x {\cal L}_{\rm m}\left(\psi^{(i)}_{\rm m}, g_{\mu\nu}^{(i)}\right)\ .
\eea
Here $\psi^{(i)}_{\rm m}$'s are matter fields, which are distinguished with each other 
by the index $(i)$. The metric tensor $g_{\mu\nu}^{(i)}$ is defined by 
\be
\label{II}
g_{\mu\nu}^{(i)}=\e^{2\beta_i\phi \over M_{\rm Pl}}g_{\mu\nu}\ .
\ee
Here $\beta_i$'s are constants, which depend on the matter. 
Typically the potential $V(\phi)$ is chosen as
\be
\label{IIIb}
V(\phi)={M^{4+n} \over \phi^n}\ .
\ee
The equation of motion for $\phi$ has the following form: 
\be
\label{III}
\nabla^2 \phi = V_{,\phi} - \sum_i {\beta_i \over M_{\rm Pl}}\e^{4\beta_i \phi \over 
M_{\rm Pl}} g^{(i)\mu\nu}T^{(i)}_{\mu\nu} \ .
\ee
If the time-development of the matter fields is small, one 
may assume $\rho_i\sim - g^{(i)\mu\nu}T^{(i)}_{\mu\nu}$. 
Here $\rho_i$'s are the energy densities of the matter fileds. 
For simplicity,  the case that  $\beta_i$'s do not depend on the 
matter fields or the case of only one kind of matter field is
considered. 
Eq.(\ref{III}) may be rewritten as
\be
\label{IV}
\nabla^2 \phi = V'(\phi) + {\beta \over M_{\rm Pl}}\e^{4\beta \phi \over 
M_{\rm Pl}} \rho_0 \ .
\ee
Let $\phi_0$ be a constant solution of (\ref{IV}), that is,
\be
\label{IVb}
0 = V'(\phi_0) + {\beta \over M_{\rm Pl}}\e^{4\beta \phi_0 \over 
M_{\rm Pl}} \rho_0 \ .
\ee
Then the effective mass of the chameleon field is given by 
\be
\label{XIII}
m^2 \equiv \left.{d \over d\phi}\left(V'(\phi_0) + {\beta \over M_{\rm Pl}}
\e^{4\beta \phi_0 \over M_{\rm Pl}} \rho_0 \right)\right|_{\phi=\phi_0}\ ,
\ee
which depends on the matter density $\rho_0$. The mass can be small at
large 
cosmological scale but might be heavy, say, on the earth. If there are $N$-matter 
fields with common $\beta$, instead of (\ref{XIII}), we have
\be
\label{XIIIb}
m^2 = \left.{d \over d\phi}\left(V'(\phi_0) + {N\beta \over M_{\rm Pl}}
\e^{4\beta \phi_0 \over M_{\rm Pl}} \rho_0 \right)\right|_{\phi=\phi_0}\ ,
\ee

Now as a quantum correction, we consider the trace anomaly due to 
$N$-conformal scalars, 
whose action is given by
\be
\label{V}
S_{\rm matter}=\sum_{i=1}^N\int d^4 x \sqrt{-g^{(i)}}\varphi^{(i)}\left(\nabla^2 - 
{1 \over 6}R^{(i)}\right)\varphi^{(i)}\ .
\ee
Here the curvature $R^{(i)}$ is constructed from the metric $g^{(i)}_{\mu\nu}$. 
In the case that the $\beta_i$ does not depend on the matter scalars, we
may write the metric as $\hat g_{\mu\nu}=g^{(i)}_{\mu\nu}
=\e^{2\beta\phi \over M_{\rm Pl}}g_{\mu\nu}$ and we express the curvature etc.
 constructed 
by $\hat g_{\mu\nu}$ as $\hat R$ etc. Then the conformal anomaly ${\cal T}$ is given by
\be
\label{VI}
{\cal T}={N \over 180\left(4\pi\right)^2}\left(\hat R_{\mu\nu\rho\sigma} 
\hat R^{\mu\nu\rho\sigma} - \hat R_{\mu\nu} \hat R^{\mu\nu} 
+ \nabla^2 \hat R\right) \ .
\ee
Since under the scale transformation $g_{\mu\nu}\to \e^\sigma g_{\mu\nu}$, 
the Riemann tensor is transformed as 
\bea
\label{VII}
R_{\zeta\mu\rho\sigma}&&\to \nn
&& \e^\sigma\left\{ R_{\zeta\mu\rho\sigma} - {1 \over 2}
\left(g_{\zeta\rho}\nabla_\nu \nabla_\mu \sigma + g_{\mu\nu}\nabla_\rho \nabla_\xi \sigma 
 - g_{\mu\rho}\nabla_\nu \nabla_\zeta \sigma - g_{\zeta\nu}\nabla_\rho \nabla_\mu \sigma 
\right) \right. \nn
&& + {1 \over 4}\left(g_{\zeta\rho}\nabla_\nu \sigma \nabla_\mu \sigma 
+ g_{\mu\nu}\nabla_\rho \sigma \nabla_\xi \sigma 
 - g_{\mu\rho}\nabla_\nu \sigma \nabla_\zeta \sigma 
 - g_{\zeta\nu}\nabla_\rho \sigma\nabla_\mu \sigma \right) \nn
&& \left. - {1 \over 4}\left(g_{\zeta\rho} g_{\mu\nu} - g_{\zeta\nu}g_{\mu\rho}\right)
\nabla_\xi \sigma \nabla^\xi \sigma \right\}\ ,
\eea
one finds that the trace anomaly can be rewritten as
(for a review of conformal anomaly for 4d dilaton coupled matter, see
\cite{SN})
\bea
\label{VIII}
{\cal T}&=& {N\e^{-2\tilde \phi} \over 180\left(4\pi\right)^2}\left[ R_{\mu\nu\rho\sigma} 
R^{\mu\nu\rho\sigma} - R_{\mu\nu} R^{\mu\nu} + \nabla^2 R \right. \nn
&& - 3\left(\nabla^2\right)^2\tilde \phi - \nabla_\mu \tilde\phi \nabla^\mu R 
 - 2 R^{\mu\nu}\nabla_\mu\nabla_\nu\tilde\phi  
+ R^{\mu\nu}\nabla_\mu \tilde\phi \nabla_\nu\tilde\phi \nn
&& -2\nabla^\mu\nabla^\nu\tilde\phi\nabla_\mu\nabla_\nu\tilde\phi 
+ 2 \left(\nabla^2\tilde\phi\right)^2 \nn
&& \left. + 2\nabla^\mu\tilde\phi\nabla^\nu\tilde\phi\nabla_\mu\nabla_\nu\tilde\phi 
+ \nabla^2\tilde\phi\nabla_\sigma\tilde\phi\nabla^\sigma\tilde\phi \right]\ .
\eea
Here
\be
\label{IX}
\tilde\phi={2\beta \over M_{\rm Pl}}\phi\ .
\ee
The quantum correction to energy density can be included by replacing 
\be
\label{X}
\rho_0\to \rho_0 - {\cal T} 
\ee
in (\ref{IV}). The next assumption is that the curvature is small and can
be 
neglected. We also consider the perturbation of
quantum corrected equation (\ref{IV}) 
by replacing 
\be
\label{XI}
\phi=\phi_0 + \delta\phi 
\ee
and keeping only the linear part of $\delta\phi$
\be
\label{XII}
\nabla^2 \delta\phi = m^2 \delta\phi + {N \beta^2 \over 30 \left(4\pi\right)^2
M_{\rm Pl}^2}\left(\nabla^2\right)^2 \delta\phi\ .
\ee
Here $m^2$ is defined by (\ref{XIIIb}). 
By replacing $\nabla^2\to \omega^2$, we obtain
\be
\label{XIV}
\omega^2 = {30 \left(4\pi\right)^2 \over N \beta^2}\left( 1 \pm 
\sqrt{ 1 - {4m^2 N \beta^2 \over 30 \left(4\pi\right)^2 M_{\rm Pl}^2}}\right)\ .
\ee
If 
\be
\label{XV}
{4m^2 N \beta^2 \over 30 \left(4\pi\right)^2 M_{\rm Pl}^2}<1\ ,
\ee
$\omega^2$ is real and positive, which indicates that the system is stable 
under the perturbations. On the other hand, if 
\be
\label{XVI}
{4m^2 N \beta^2 \over 30 \left(4\pi\right)^2 M_{\rm Pl}^2}>1\ ,
\ee
$\omega^2$ becomes complex. Then the system becomes instable.  
If the matter density is large $\phi_0\sim 0$, 
we find ${nM^{4+n} \over \phi_0^{n+1}}\sim {\beta \rho_0 \over M_{\rm Pl}}$ 
from (\ref{IVb}) for the potential $V(\phi)=M^{4+n}\phi^{-n}$. 
Then we have
\be
\label{AA1}
m^2={n(n+1) M^{4+n} \over \phi_0^{n+2}} + {4\beta^2 \rho_0 \over M_{\rm Pl}^2}
\e^{4\beta\phi_0 \over M_{\rm Pl}}\sim M^{-{n+4 \over n+1}}
M_{\rm Pl}^{-{n+2 \over n+1}}\rho_0^{n+2 \over n+1}\ .
\ee
Then the condition (\ref{XVI}) can be rewritten as
\be
\label{XIVb}
\rho_0 \gtrsim M^{n+4 \over n+2}M_{\rm Pl}^{3n+4 \over n+2}\ .
\ee
For $n\to\infty$, we have $\rho_0\gtrsim MM_{\rm Pl}^3$ or $M\lesssim M_{\rm Pl}\left({\rho \over 
M_{\rm Pl}^4}\right)$. On the other hand, for $n\to 0$ (if we include the case that 
$n$ is fractional or irrational), we have $\rho_0\gtrsim M^2 M_{\rm Pl}^2$ or 
$M\lesssim M_{\rm Pl}\sqrt{\rho \over M_{\rm Pl}^4}$. We should note 
$M_{\rm Pl}\sim \left(10^{-35}{\rm m}\right)^{-1}\sim 10^{19}$GeV$\sim 10^{-5}$g, 
or $M_{\rm Pl}^4\sim 10^{94}$g/cm$^3$. 
In case of white dwarf, we have $\rho\sim 10^6$g/cm$^3$, therefore, if 
$M\lesssim 10^{-60}$eV (for $n\to \infty$) or $M\lesssim 10^{-16}$eV (for $n\to 0$), 
there might be an instability.  
On the other hand, in case of neutron star, we have $\rho\sim 10^{14}$g/cm$^3$. 
Then if $M\lesssim 10^{-52}$eV (for $n\to \infty$) or $M\lesssim 10^{-12}$eV (for $n\to 0$), 
there might be an instability.  
Such values of $M$ seem to be unnatural. 
Therefore, chameleon  cosmology (at least, in
newtonian limit \cite{KW}) seems to be 
stable under the perturbations with the account of quantum effects.
Nevertheless, further checks of the consistency of chameleon cosmology 
should be fulfilled.

Eq.(\ref{XVI}) seems to indicate that if $m^2$ is very large, there might
appear an instability. 
In such a case, one cannot neglect the curvature. 
In the following,  the case that the curvature is not small but (covariantly) 
constant is considered. The Riemann curvature can be written as 
\be
\label{XVII}
R_{\mu\nu\rho\sigma}={1 \over l^2}\left(g_{\mu\rho}g_{\nu\sigma} 
 - g_{\mu\sigma}g_{\nu\rho}\right)\ .
\ee
As we will see later, the mode cooresponding to the scale of the metric is
mixed with 
the chameleon scalar  in the perturbation. Before going to the
chameleon 
theory with quantum correction, we consider the perturbations in usual
Einstein gravity. 

Multiplying $g^{\mu\nu}$ to the Einstein equation
\be
\label{Q1}
R_{\mu\nu} - {1 \over 2}g_{\mu\nu}R={\Lambda \over 2}g_{\mu\nu}\ ,
\ee
one obtains (for 4-dim. case, for simplicity)
\be
\label{Q2}
- R=2\Lambda\ .
\ee
For the general variation of the metric
\be
\label{Q3}
g_{\mu\nu}\to g_{\mu\nu} + \delta g_{\mu\nu}\ ,
\ee
we have 
\be
\label{Q4}
\delta R = -\delta g_{\mu\nu} R^{\mu\nu} + \nabla^\mu \nabla^\nu \delta g_{\mu\nu} 
 - \nabla^2 \left(g^{\mu\nu}\delta g_{\mu\nu}\right)\ .
\ee
Choosing a gauge condition
\be
\label{Q5}
\nabla^\mu \delta g_{\mu\nu}=0\ ,
\ee
and considering the perturbations on the deSitter space in
(\ref{XVII}), 
we obtain, from the reduced Einstein equation (\ref{Q2}), 
\be
\label{Q7}
0=-{3 \over l^2}\delta G - \nabla^2 \delta G\ .
\ee
Here
\be
\label{Q8}
\delta G = g^{\mu\nu}\delta g_{\mu\nu}\ ,
\ee
which corresponds to the mode of the scale of the metric tensor. 
Eq.(\ref{Q7}) seems to indicate that in above gauge $\delta G$ has a
tachyonic mass,
\be
\label{Q9}
m_T^2 = - {3 \over l^2}\ .
\ee

In case of the chameleon theory with  quantum effects, the equation
corresponding to 
the reduced Einstein equation (\ref{Q2}) has the following form:
\be
\label{XVIII}
-{M_{\rm Pl}^2 \over 2}R + {1 \over 2}\partial_\mu \phi \partial^\mu \phi + 2V(\phi) 
={1 \over 2}\e^{4\beta\phi \over M_{\rm Pl}}\left(-N\rho_0 + {\cal T}\right)\ .
\ee
It is	 assumed again that there are $N$ matter fields (conformal
scalars)
with a common $\beta_i=\beta$. For covariantly constant curvature 
as in (\ref{XVII}) and constant chameleon field, $\phi=\phi_0$,
Eq.(\ref{XVIII}) 
reduces as
\be
\label{XIX}
 - {6M_{\rm Pl}^2 \over l^2} + 2V(\phi_0)=-{1 \over 2}N\rho_0\e^{4\beta\phi_0 \over M_{\rm Pl}}
 - {N \over 30\left(4\pi\right)^2 l^4}\ .
\ee
On the other hand, Eq.(\ref{IV}) after the replacement (\ref{X}) has the following 
form:
\be
\label{XX}
0=V'(\phi_0) + {N\beta\rho_0 \over M_{\rm Pl}}\e^{4\beta\phi_0 \over M_{\rm Pl}} 
+ {N\beta \over 15\left(4\pi\right)^2 l^4 M_{\rm Pl}}\ .
\ee
Combining (\ref{XIX}) with (\ref{XX}),  $l^2$ and $\phi_0$ may be
evaluated. 
The next is to address the perturbations around the solution. 
For the perturbation, we choose a gauge condition (\ref{Q5}). Then 
\bea
\label{XXI}
\delta{\cal T}&=&{N\e^{-{4\beta\phi_0 \over M_{\rm Pl}}} \over 180\left(4\pi\right)^2}
\left[{4\beta \over M_{\rm Pl}}\left({12 \over l^4}\delta\phi - {24 \over l^2}\nabla^2 
\delta\phi - 3\left(\nabla^2\right)^2\delta\phi\right) \right. \nn
&& \left. + {6 \over l^4}\delta G - {1 \over l^2}\nabla^2 \delta G - \left(\nabla^2\right)^2
\delta G \right]\ .
\eea
Here $\delta G$ is defined by (\ref{Q8}). The equation (\ref{XVIII}) gives
\bea
\label{XXII}
&& - {M_{\rm Pl} \over 2}\left(- {3 \over l^2}\delta G - \nabla^2 \delta G\right) 
+ 2V'(\phi_0)\delta \phi \nn
&& = - {2N\beta\rho_0 \over M_{\rm Pl}}\e^{4\beta\phi_0 \over M_{\rm Pl}}\delta\phi 
+ {N \over 360\left(4\pi\right)^2} \left[{4\beta \over M_{\rm Pl}}\left(- {24 \over l^2}\nabla^2 
\delta\phi - 3\left(\nabla^2\right)^2\delta\phi\right) \right. \nn
&& \quad \left. + {6 \over l^4}\delta G - {1 \over l^2}\nabla^2 \delta G
 - \left(\nabla^2\right)^2 \delta G \right]\ .
\eea
On the other hand, Eq.(\ref{IV}) after the replacement (\ref{X}) looks
like
\bea
\label{XXIII}
\nabla^2 \delta\phi &=& m^2 \delta\phi - {N\beta \over 180\left(4\pi\right)^2 M_{\rm Pl}} 
\left[{4\beta \over M_{\rm Pl}}\left(- {24 \over l^2}\nabla^2 
\delta\phi - 3\left(\nabla^2\right)^2\delta\phi\right) \right. \nn
&& \quad \left. + {6 \over l^4}\delta G - {1 \over l^2}\nabla^2 \delta G
 - \left(\nabla^2\right)^2 \delta G \right]\ .
\eea
Here $m^2$ is defined by (\ref{XIIIb}). Defining 
\be
\label{XXIV}
b\equiv {N \over 120\left(4\pi\right)^2}\ ,
\ee 
and  replacing $\nabla^2\to \omega^2$, we obtain
\bea
\label{XXV}
0&=&{4\beta b \over M_{\rm Pl}}\left(\omega^4 + {8 \over l^2}\omega^2 + {4 \over l^4}\right)
\delta\phi \nn
&& + \left(\omega^2 + {3 \over l^2}\right)\left({b \over 3l^2}\omega^2 
 - {2b \over 3l^2} + {M_{\rm Pl}^2 \over 2}\right)\delta G\ ,\\
\label{XXVI}
0&=& \left\{ {8\beta^2 b \over M_{\rm Pl}^2}\omega^4 + \left({64\beta^2 b \over 
M_{\rm Pl}^2 l^2} - 1 \right)\omega^2 + m^2 \right\}\delta\phi \nn
&& + {2\beta b \over 3M_{\rm Pl}}\left(\omega^2 + {3 \over l^2}\right)
\left(\omega^2 - {2 \over l^2}\right) \delta G\ .
\eea
Here Eq.(\ref{XX}) is used. 
In order that Eqs.(\ref{XXV}) and (\ref{XXVI}) have a non-trivial solution, the following 
condition should be satisfied,
\bea
\label{XXVII}
0&=&D(\omega^2) \nn
&\equiv& \left(\omega^2 + {3 \over l^2}\right)\left\{ {b \over 3}\left(1 - 12\beta^2\right)
\omega^4 + \left({M_{\rm Pl}^2 \over 2} - {2b \over 3l^2}\right)\omega^2 \right.\nn
&& \left. - {64 b \beta \over M_{\rm Pl}l^6} - \left({M_{\rm Pl}^2 \over 2}
 - {2b \over 3l^2}\right)
m^2 \right\}\ .
\eea
The solutions of the above equation (\ref{XXVII}) with respect to $\omega^2$ are given by
\bea
\label{XXVIII}
&& \omega^2 = - {3 \over l^2}\ , \\
\label{XXIX}
&& \omega^2 = {3 \over 2b \left(12\beta^2 - 1 \right)}\left[
{M_{\rm Pl}^2 \over 2} - {2b \over 3l^2} \right. \\
&& \left. \pm \sqrt{
\left({M_{\rm Pl}^2 \over 2} - {2b \over 3l^2}\right)^2 -{4b\left(12\beta^2 - 1 \right) \over 3}
\left\{{64 b \beta \over M_{\rm Pl}l^6} + \left({M_{\rm Pl}^2 \over 2} - {2b \over 3l^2}\right)
m^2 \right\} } \right]\ . \nonumber
\eea
The solution (\ref{XXVIII}) corresponds to the tachyonic mass in (\ref{Q9}). 
Then we neglect the solution (\ref{XXVIII}). 
We now consider the case that the curvature is  smaller than the Planck scale 
${1 \over l^2}\ll M_{\rm Pl}^2$. Then one may obtain 
${M_{\rm Pl}^2 \over 2} - {2b \over 3l^2}>0$. If $\beta^2>{1 \over 12}$,
both of the 
solutions  (\ref{XXIX}) are positive, and there is no instability. On the
other hand 
if $\beta^2<{1 \over 12}$, one of the solutions in (\ref{XXIX}) is negative, then the system 
becomes instable. 

In summary, our study indicates that quantum effects related with
conformal anomaly introduce potentially dangerous higher derivative terms 
to the equations of motion for chameleon fields. Such quantum effects are
known to lead to
so-called trace anomaly driven inflation \cite{star} at the early
universe. (For extension of anomaly driven inflation for scalar-tensor 
theories, see \cite{sergio}).
However, for chameleon cosmology in the limit of small
curvature\cite{KW} the induced instabilities are negligible.
Nevertheless, with the increase of the curvature such instabilities 
are getting more important and may put some extra limits to chameleon
cosmology (if it will be realized as realistic cosmology).

\

\noindent
{\bf Acknowledgments} 
We thank E. Elizalde,  J. Khoury, J. Lidsey and A. Starobinsky 
for helpful
discussions.
The research is supported in part by the Ministry of
Education, Science, Sports and Culture of Japan under the grant n.13135208
(S.N.),  RFBR grant 03-01-00105
(S.D.O.) and LRSS grant 1252.2003.2 (S.D.O.).

\end{document}